\documentclass[conference]{IEEEtran}
\bstctlcite{BSTcontrol}
\IEEEoverridecommandlockouts
\usepackage{cite}
\usepackage{amsmath,amssymb,amsfonts}
\usepackage{algorithmic}
\usepackage[T1]{fontenc}
\usepackage{xcolor}          
\usepackage{tikz}            
\usetikzlibrary{shapes, arrows.meta, positioning, shadows} 
\usepackage{graphicx}   
\usepackage{textcomp}
\usepackage[hidelinks]{hyperref}

\def\BibTeX{{\rm B\kern-.05em{\sc i\kern-.025em b}\kern-.08em
    T\kern-.1667em\lower.7ex\hbox{E}\kern-.125emX}}
\begin{document}

\title{``Your Doctor is Spying on You'': An Analysis of Data Practices in Mobile Healthcare Applications}

\author{\IEEEauthorblockN{Luke Stevenson}
\IEEEauthorblockA{\textit{Computer Science} \\
\textit{University of Denver}\\
Denver, Colorado, USA.\\
luke.stevenson@du.edu}
\and
\IEEEauthorblockN{Sanchari Das}
\IEEEauthorblockA{\textit{Information Sciences and Technology} \\
\textit{George Mason University}\\
Fairfax, Virginia, USA. \\
sdas35@gmu.edu}
}

\maketitle

\begin{abstract}
Mobile healthcare (\textit{mHealth}) applications promise convenient, continuous patient–provider interaction but also introduce severe and often underexamined security and privacy risks. We present an end-to-end audit of $272$ Android mHealth apps from Google Play, combining \textit{permission forensics}, \textit{static vulnerability analysis}, and \textit{user review mining}. Our multi-tool assessment with MobSF, RiskInDroid, and OWASP Mobile Audit revealed systemic weaknesses: $26.1\%$ request fine-grained location without disclosure, $18.3\%$ initiate calls silently, and $73$ send SMS without notice. Nearly half ($49.3\%$) still use deprecated \texttt{SHA-1} encryption, $42$ transmit unencrypted data, and $6$ remain vulnerable to \emph{StrandHogg~2.0}. Analysis of \(2.56\text{M}\) user reviews found $28.5\%$ negative or neutral sentiment, with over $553,000$ explicitly citing privacy intrusions, data misuse, or operational instability. These findings demonstrate the urgent need for enforceable permission transparency, automated pre-market security vetting, and systematic adoption of secure-by-design practices to protect Protected Health Information (PHI).
\end{abstract}


\section{Introduction}
\label{sec:intro}
Mobile healthcare (\textit{mHealth}) applications are reshaping digital health by enabling continuous, remote patient–provider interaction across diagnostics, treatment, and monitoring workflows. They deliver capabilities ranging from real-time vital sign monitoring and medication adherence tracking to teleconsultations and insurance claims management~\cite{mosa2012systematic,4278593}. However, their deep integration with mobile devices and direct handling of Protected Health Information (PHI) greatly expands the attack surface for privacy breaches and malicious exploitation~\cite{duckert2022protecting,tazi2024evaluating,saka2024evaluating,das2022privacy,tazi2024sok,tazi2022sok}. Breaches of PHI can cause severe and long-lasting harm, including large-scale identity theft, fraudulent insurance claims, targeted discrimination, and psychological distress~\cite{6906856,osawaru2024electronic,surani2023security,surani2022understanding}. Regulatory frameworks such as HIPAA~\cite{6845198} and GDPR~\cite{tzanou2023health} mandate strict controls over data collection, storage, transmission, and sharing. However, enforcement remains inconsistent, particularly for consumer-oriented mHealth applications that fall outside medical device classifications~\cite{grover2025sok,adhikari2025natural,tazi2024we,adhikari2023evolution,tazi2024improving,monroe2021location}. This regulatory blind spot enables insecure default configurations, deprecated cryptographic protocols, and opaque data-sharing practices to proliferate in app marketplaces with minimal oversight.

To examine these risks, we conducted a multi-faceted audit of $272$ Android mHealth applications from Google Play. Using \textit{MobSF}, \textit{RiskInDroid}, and \textit{OWASP Mobile Audit}, we performed static vulnerability analysis, permission forensics, and cryptographic inspections. These technical assessments were paired with sentiment and thematic analysis of over $2.5M$ user reviews, enabling correlation between measurable security weaknesses and user-reported trust, privacy, and usability concerns. This study addresses three research questions:

\noindent \textbf{\textit{RQ1:}} What proportion of mHealth applications exhibit permission overreach, insecure cryptographic implementations, or exploitable vulnerabilities, and how are these weaknesses characterised in static analysis?

\noindent \textbf{\textit{RQ2:}} To what extent do technical weaknesses such as insecure permission use, outdated cryptography, and misconfigured network security correlate with user-reported privacy, security, and usability issues in mHealth applications?

\noindent \textbf{\textit{RQ3:}} What misconfiguration and permission misuse patterns can be extracted from this evidence, and how can they inform enforceable secure-by-design and regulatory practices?

Our evaluation uncovered a layered security and privacy deficit across the mHealth ecosystem. Manifest inspection revealed $2{,}252$ instances of unnecessarily exported broadcast receivers and $1{,}232$ cases of exported permissions without defined protection levels. Certificate analysis showed $22$ applications configured to trust all TLS certificates and $2$ with certificate pinning deliberately disabled. RiskInDroid detected frequent misuse of high-impact capabilities, including \texttt{WRITE\_SETTINGS} in $130$ applications and \texttt{MANAGE\_ACCOUNTS} in $117$, often without any declared need. OWASP Mobile Audit reported an average of $44$ critical and more than $2{,}000$ high-severity issues per app, with improper credential handling and insecure communication among the most common weaknesses. Review mining revealed that usable security failures, including broken authentication flows and inconsistent password policies between mobile and web clients, accounted for $553,495$ reviews, representing $21.6\%$ of all feedback and exceeding direct privacy-related complaints. These results show that mHealth risks are deeply embedded in both code-level design choices and operational execution.

\begin{itemize}
\item We conduct an end-to-end security evaluation of $272$ Android mHealth applications, integrating static bytecode and manifest auditing, fine-grained permission usage forensics from API call tracing, and sentiment and topic mining of \(2.56\text{M}\) user reviews to build a multi-dimensional risk model.
\item We uncover exploitable weaknesses including undeclared and over-privileged permissions, invocation of proprietary and undocumented APIs, reliance on deprecated cryptographic primitives such as \texttt{RSA~SHA-1}, misconfigured network security policies allowing cleartext traffic, and insecure $X.509$ certificate validation bypasses enabling man-in-the-middle interception of PHI.
\item We identify statistically significant correlations between technical weaknesses and user sentiment: apps exhibiting permission overreach (\emph{$r = 0.62, p < 0.001$}), insecure cryptography (\emph{$r = 0.54, p < 0.001$}), or unencrypted data transmission (\emph{$r = 0.47, p < 0.001$}) show substantially higher proportions of negative or neutral reviews (\emph{$28.5\%$} overall). Notably, \emph{$553,495$} reviews explicitly mention privacy intrusions, data misuse, or operational instability, underscoring the link between vulnerabilities and trust.
\item We identify systemic failure patterns across the mHealth ecosystem, including persistent non-compliance with least-privilege principles, absence of certificate pinning, and inadequate enforcement of secure transport protocols. We translate these into concrete, enforceable recommendations for developer toolchains, app marketplace vetting workflows, and regulatory audits to embed secure-by-design principles prior to deployment.
\end{itemize}

\section{Background}
\label{sec:background}
Security and privacy in mobile applications are critical, particularly in open ecosystems like Android, where platforms such as Google Play allow unrestricted app publication~\cite{das2024design,saka2024evaluating,kishnani2023assessing,kishnani2024towards,kishnani2024dual,kishnani2022privacy,neupane2022data}. While this openness fosters innovation, it also introduces risks of security flaws, privacy breaches, and regulatory noncompliance~\cite{taylor2016quantifying}. These risks are amplified in healthcare apps that handle PHI and are subject to strict legal and ethical standards. Sun et al.~\cite{sun2020oat} studied attack detection and integrity in healthcare embedded systems. Other studies have examined runtime behaviors, usability, and permission misuse~\cite{marky2020don,taylor2016quantifying}, yet none have simultaneously considered undeclared permissions, cryptographic weaknesses, and user concerns in healthcare. This study addresses that gap.

\subsection{Tools}
Over the past decade, open-source and academic communities have developed mature frameworks for static and dynamic mobile app analysis, enabling the detection of insecure permissions, weak cryptography, and privacy-invasive behaviors beyond superficial inspection. We employ three complementary tools in this study. The \emph{Mobile Security Framework (MobSF)}~\cite{bergadano2020modular} performs static APK analysis to identify vulnerabilities such as weak hash functions, improper intent handling, cleartext traffic, and excessive permission requests, assigning severity scores for comparative profiling~\cite{tileria2020wearflow,kalhor2023evaluating,jensen2021multi,noah2024aging,noah2024evaluating,noah2022privacy}. \emph{RiskInDroid}~\cite{merlo2017riskindroid} extends permission analysis by categorizing permissions into declared, used, undeclared but used, and required but unused~\cite{khorkheli2024improving}. This approach is effective for detecting stealthy access to sensitive APIs in healthcare contexts. The \emph{OWASP Mobile Audit}\cite{acharya2015owasp}, aligned with the OWASP Mobile Application Security Testing Guide (MASTG)\cite{holguera2022owasp}, applies Static Application Security Testing (SAST) to uncover vulnerabilities ranked from critical to low severity~\cite{kishnani2024securing}. We also reference auxiliary tools such as APKLeaks~\cite{apkleaks2021} and AndroBugs~\cite{AndroBugs2023framework} for lightweight scanning, and note that hybrid fuzzing and penetration testing~\cite{pandey2021web} expand methodological coverage. These capabilities are integrated into a reproducible pipeline tailored for mHealth, targeting code-level threats and privacy practices that circumvent platform enforcement.

\subsection{Mobile Application Privacy and Security}
Mobile applications frequently request access to sensitive data and services beyond what is functionally necessary~\cite{leontiadis2012don,momenzadeh2020bayesian,das2020user,tazi2023accessibility,sun2016blender,momenzadeh2021bayesian,podapati2025sok}. While Android's permission model is designed to promote transparency, users often approve requests without understanding their implications~\cite{fife2012privacy,streiff2019overpowered,debnath2020studies,dev2019personalized}. As a result, permissions may be legally obtained but ethically questionable, particularly when purposes are obscured or undeclared. Prior studies~\cite{10.1145/2420950.2420956,tvidas2011all,sivan2019analysis} report persistent requests for access to location, camera, SMS, and contact data without clear justification. Liu et al.~\cite{LIU20112022} found that developers frequently violate the principle of least privilege, requesting access to sensitive resources without functional necessity. Our findings confirm and extend these trends in the mHealth domain. Zhu et al.~\cite{10.1145/2623330.2623705} advocate enforceable, user-centric privacy frameworks, yet enforcement under GDPR and HIPAA remains inconsistent, particularly for consumer-facing health apps that are not subject to medical device regulation~\cite{carayon2018challenges}. Our findings highlight this gap between regulatory intent and practical enforcement, demonstrating that even highly popular healthcare apps bypass platform protections and fail to meet privacy-by-design standards.

\subsection{Healthcare Applications}
mHealth applications manage PHI such as clinical histories, biometric readings, medication records, and behavioral data, making data confidentiality and integrity both technical and ethical imperatives. Prasanna et al.\cite{prasanna2013decision} classified mHealth into domains ranging from disease management to patient education, revealing highly variable risk profiles. This diversity complicates the uniform application of security standards. Regulatory fragmentation exacerbates the issue. In the United States, only apps that replace or augment clinical devices fall under HIPAA and FDA oversight\cite{LIU20112022}, leaving wellness and lifestyle apps largely unregulated~\cite{moen2005health}. Reddy et al.~\cite{reddy2023review} documented major breaches linked to insecure healthcare apps, often caused by outdated encryption, absent access controls, or weak transport-layer security. Park and Youm~\cite{park2022appsecurity} and Ahmad et al.~\cite{ahmad2023leveraging} similarly report persistent use of deprecated algorithms, insecure data storage, and excessive permission requests. An additional risk arises from data brokerage. Shah and Lim~\cite{shah2024analysis} found that many mHealth apps embed advertising SDKs and analytics services that exfiltrate user data without meaningful consent, effectively monetizing sensitive health information\cite{fernandez2018analytic}. While GDPR and HIPAA establish strong principles, enforcement remains inconsistent. Liu et al.\cite{LIU20112022} call for extending regulation to all apps handling PHI, regardless of classification. Amin et al.\cite{amin2024balancing} explore how to balance privacy preservation with secure data handling in compliance contexts. Our study addresses these blind spots through a fine-grained, tool-assisted analysis of $272$ Android healthcare apps, identifying permission abuses, reliance on legacy cryptography, and insecure certificate handling. By triangulating technical findings with over $2.5$M user reviews, we reveal a consistent pattern in which usability failures and opaque data practices undermine trust in digital healthcare.
\section{Method}
\label{sec:method}

Our methodology consisted of three phases: (1) Data Collection and Preprocessing, (2) Security and Privacy Analysis, and (3) User Experience Analysis, as shown in Figure~\ref{fig:study_design}.

\definecolor{lightorange}{RGB}{253, 227, 203}
\definecolor{lightgreen}{RGB}{217, 242, 217}
\definecolor{lightgray}{RGB}{235, 235, 235}
\definecolor{lightred}{RGB}{250, 218, 221}
\definecolor{lightcyan}{RGB}{213, 236, 248}

\tikzstyle{phase} = [rectangle, rounded corners, minimum width=4cm, minimum height=1cm, text centered, draw=black, fill=lightorange, font=\small]
\tikzstyle{reviewpositive} = [rectangle, minimum width=3cm, minimum height=1cm, text centered, draw=black, fill=lightgreen, font=\small]
\tikzstyle{reviewneutral} = [rectangle, minimum width=3cm, minimum height=1cm, text centered, draw=black, fill=lightgray, font=\small]
\tikzstyle{reviewnegative} = [rectangle, minimum width=3cm, minimum height=1cm, text centered, draw=black, fill=lightred, font=\small]
\tikzstyle{tool} = [rectangle, minimum width=3cm, minimum height=1cm, text centered, draw=black, fill=lightcyan, font=\small, drop shadow]
\tikzstyle{arrow} = [thick,->,>=stealth]

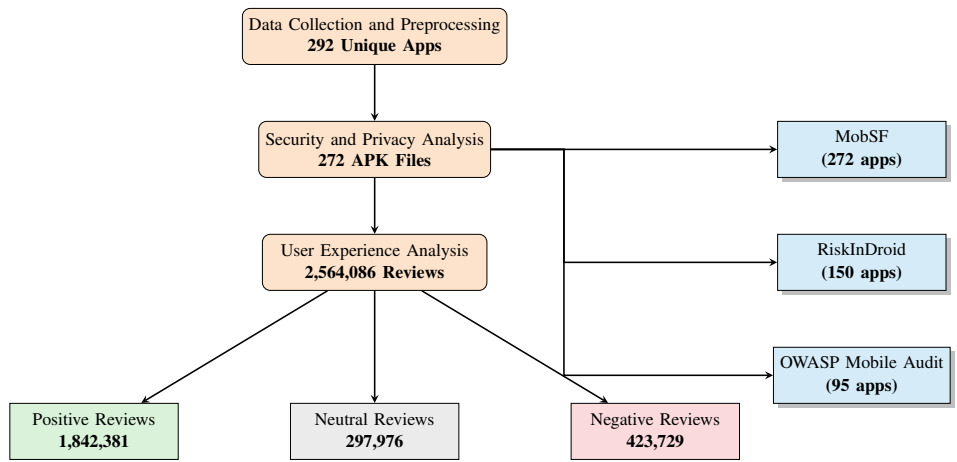
\begin{figure*}[t]
    \centering
    \resizebox{0.7\linewidth}{!}{ 
    \begin{tikzpicture}[node distance=1cm and 1cm]
\node (data) [phase, align=center] {Data Collection and Preprocessing \\[0.1em] \textbf{292 Unique Apps}};
\node (security) [phase, below=of data, align=center] {Security and Privacy Analysis \\[0.1em] \textbf{272 APK Files}};
\node (user) [phase, below=of security, align=center] {User Experience Analysis \\[0.1em] \textbf{2,564,086 Reviews}};

     \node (positive) [reviewpositive, below=of user, xshift=-5cm, yshift=-1cm, align=center] {Positive Reviews \\[0.1em] \textbf{1,842,381}};
     \node (neutral)  [reviewneutral, below=of user, yshift=-1cm, align=center] {Neutral Reviews \\[0.1em] \textbf{297,976}};
     \node (negative) [reviewnegative, below=of user, xshift=5cm, yshift=-1cm, align=center] {Negative Reviews \\[0.1em] \textbf{423,729}};

\node (mobsf) [tool, right=5.1cm of security, align=center] {MobSF \\[0.2em] \textbf{(272 apps)}};
\node (risk) [tool, below=of mobsf, align=center] {RiskInDroid \\[0.2em] \textbf{(150 apps)}};
\node (owasp) [tool, below=of risk, align=center] {OWASP Mobile Audit \\[0.2em] \textbf{(95 apps)}};

        \draw [arrow] (data) -- (security);
        \draw [arrow] (security) -- (user);

        \draw [arrow] (user) -- (positive);
        \draw [arrow] (user) -- (neutral);
        \draw [arrow] (user) -- (negative);

        \draw [arrow] (security.east) -- ++(1.3,0) |- (mobsf.west);
        \draw [arrow] (security.east) -- ++(1.3,0) |- (risk.west);
        \draw [arrow] (security.east) -- ++(1.3,0) |- (owasp.west);
    \end{tikzpicture}
    }
    \caption{Study design showing data collection, static security analysis (MobSF, RiskInDroid, OWASP), and sentiment-based review analysis.}
    \label{fig:study_design}
\end{figure*}

We identified healthcare-related Android applications from Google Play using twenty targeted search queries such as \emph{healthcare}, \emph{medical}, \emph{fitness tracker}, and \emph{telemedicine}. The \texttt{google-playstore-scraper} library retrieved the top thirty ranked apps per query, capturing metadata including app IDs, install ranges, developer names, and last update timestamps. Duplicate detection was performed using $SHA$-$256$ hashing of package names and version codes, yielding $292$ unique applications. APKs were retrieved from Evozi and APKPure to capture distributable binaries, with preference for the latest release matching the Play Store version. File integrity was verified through cryptographic hash matching, and incomplete or corrupted downloads were discarded. This resulted in $272$ valid APKs for static security assessment.

The security and privacy analysis combined three complementary static frameworks to maximise coverage and cross-validation of findings.~\emph{MobSF} analyzed all $272$ APKs in static mode, disassembling bytecode via \texttt{baksmali} and scanning for insecure network endpoints, weak cryptographic primitives, outdated certificate chains, exposed activities or services, and misconfigured manifest entries. Detected issues were assigned severity levels aligned with CWE identifiers, enabling systematic comparison with known vulnerability classes.~\emph{RiskInDroid} processed $150$ APKs to generate a mapping between requested permissions in the manifest and observed API calls in the compiled code, identifying undeclared-but-used permissions as potential covert channels for sensitive data access. Proprietary and undocumented permissions were flagged by comparing extracted entries against the Android \texttt{android.Manifest.permission} namespace, and cross-checked against MobSF's API call traces for corroboration.~\emph{OWASP Mobile Audit} examined $95$ APKs using static application security testing rules, detecting insecure storage practices such as unencrypted local files or shared preference misuse, hardcoded credentials within source code, and insufficient input validation. Each finding was classified under the OWASP Mobile Top 10 categories to aid severity prioritization and alignment with industry benchmarks.

For user experience analysis, we collected $2,564,086$ reviews from Google Play, including review text, star ratings, app version, timestamp, and user locale. Preprocessing included Unicode normalization, tokenization, stopword removal, and lemmatization to ensure linguistic consistency across English and non-English content. Sentiment analysis was performed using \textit{TextBlob}, applying polarity thresholds of $>0.1$ for positive, between $-0.1$ and $0.1$ for neutral, and $<-0.1$ for negative sentiment. This yielded 1,842,381 positive, $297,976$ neutral, and $423,729$ negative reviews. The combined $730,705$ neutral and negative entries were subjected to n-gram frequency modelling and co-occurrence network analysis to identify patterns of terms associated with security, privacy, and functionality complaints. Manual thematic coding classified feedback into six domains: privacy, security, performance, usability, data collection, and functionality. These themes were cross-referenced with static analysis results to correlate perceived trust issues with objective security weaknesses.

\section{Results}
\label{sec:results}

\subsection{MobSF}
MobSF assigned a security score to each application based on high- and medium-severity findings across permissions, network security, certificate handling, manifest configuration, and code base analysis. In our dataset, scores ranged from $35$ (least secure) to $60$ (most secure), with two applications receiving a grade of C. The average score across all $272$ applications was $47$, indicating that security weaknesses are widespread rather than isolated anomalies.

Dangerous permissions were both common and diverse. Frequently requested high-risk permissions included \texttt{POST\_NOTIFICATIONS} in $221$ applications, \texttt{CAMERA} in $176$, \texttt{WRITE\_EXTERNAL\_STORAGE} in $204$, \texttt{READ\_EXTERNAL\_STORAGE} in $188$, and \texttt{ACCESS\_FINE\_LOCATION} in $146$. In total, MobSF detected $529$ proprietary or undocumented permissions, often introduced through third-party SDKs. These included capabilities to access all downloads, read unique device identifiers, and write directly to internal storage. Because such permissions lie outside Android's standard auditing model, their behavior is concealed from both end-users and independent security reviewers.

Network security analysis revealed severe misconfigurations. Twenty-two applications were configured to trust all TLS certificates, $42$ allowed cleartext HTTP traffic, and two explicitly disabled certificate pinning. Each of these issues exposes PHI to interception on unsecured or malicious networks. Cryptographic inspection identified $134$ applications vulnerable to the \emph{Janus} exploit through flawed signature schemes, $58$ using \texttt{SHA1-RSA} despite documented collision vulnerabilities, and $9$ relying on \texttt{MD5}, a hash function that is considered fully broken.

\begin{figure*}[t]
    \centering
    \includegraphics[width=0.95\linewidth]{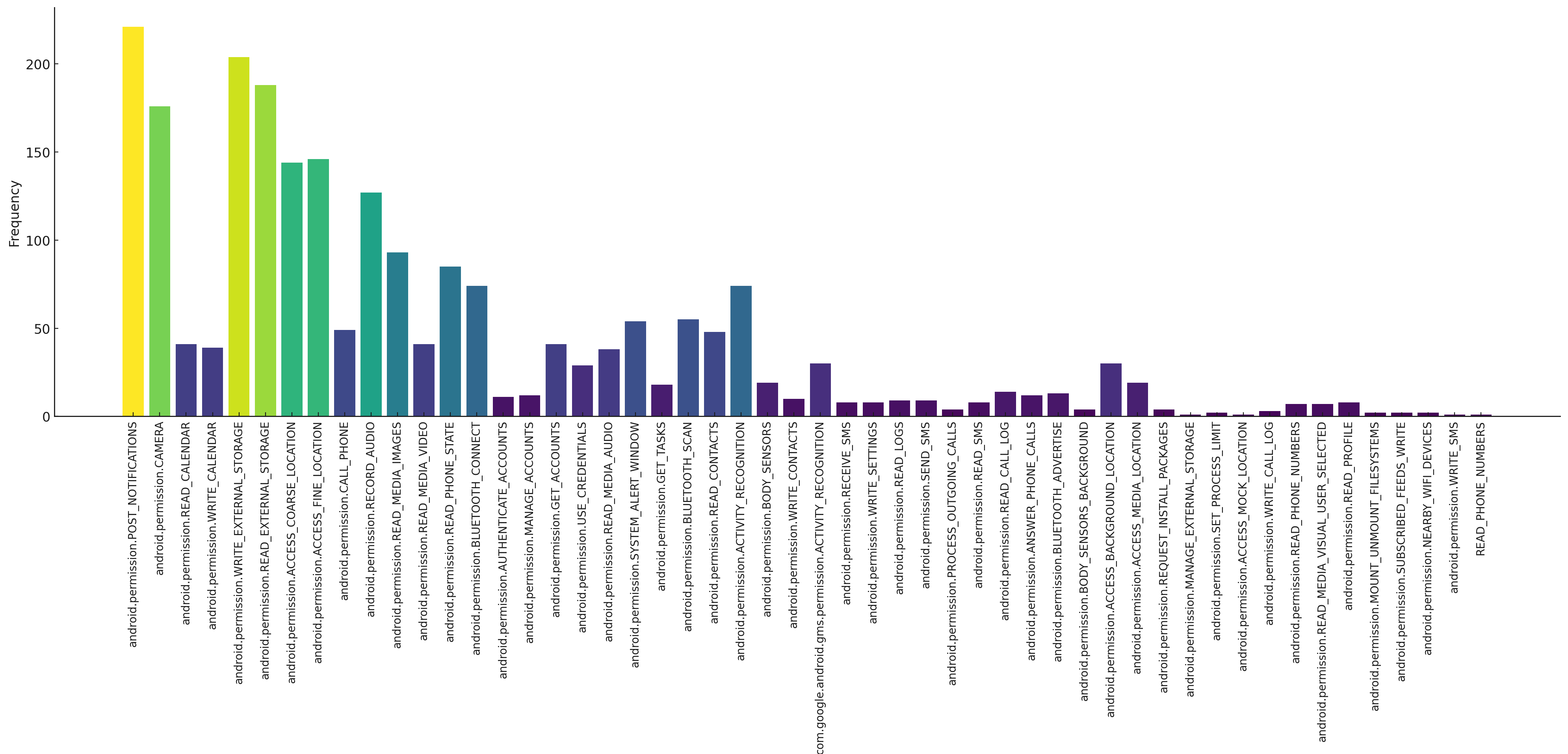}
    \caption{Distribution of dangerous permissions identified across healthcare applications using RiskInDroid.}
    \label{fig:dangerousmob}
\end{figure*}

Manifest-level analysis revealed several systemic risks. High-severity findings included 181 apps installable on outdated Android versions, $118$ exposing public asset links, and $6$ vulnerable to task hijacking via \emph{StrandHogg~$2.0$}. Frequent issues involved $2,252$ cases of \texttt{android:exported="true"} exposing broadcast receivers, $1,232$ exported permissions without protection levels, and $321$ uses of \texttt{taskAffinity} that could enable intent interception. Figure~\ref{fig:dangerousmob} shows the distribution of dangerous permissions, emphasizing widespread access to location, camera, and external storage.

\subsection{RiskInDroid}
\begin{figure*}[t]
    \centering
    \includegraphics[width=0.95\linewidth]{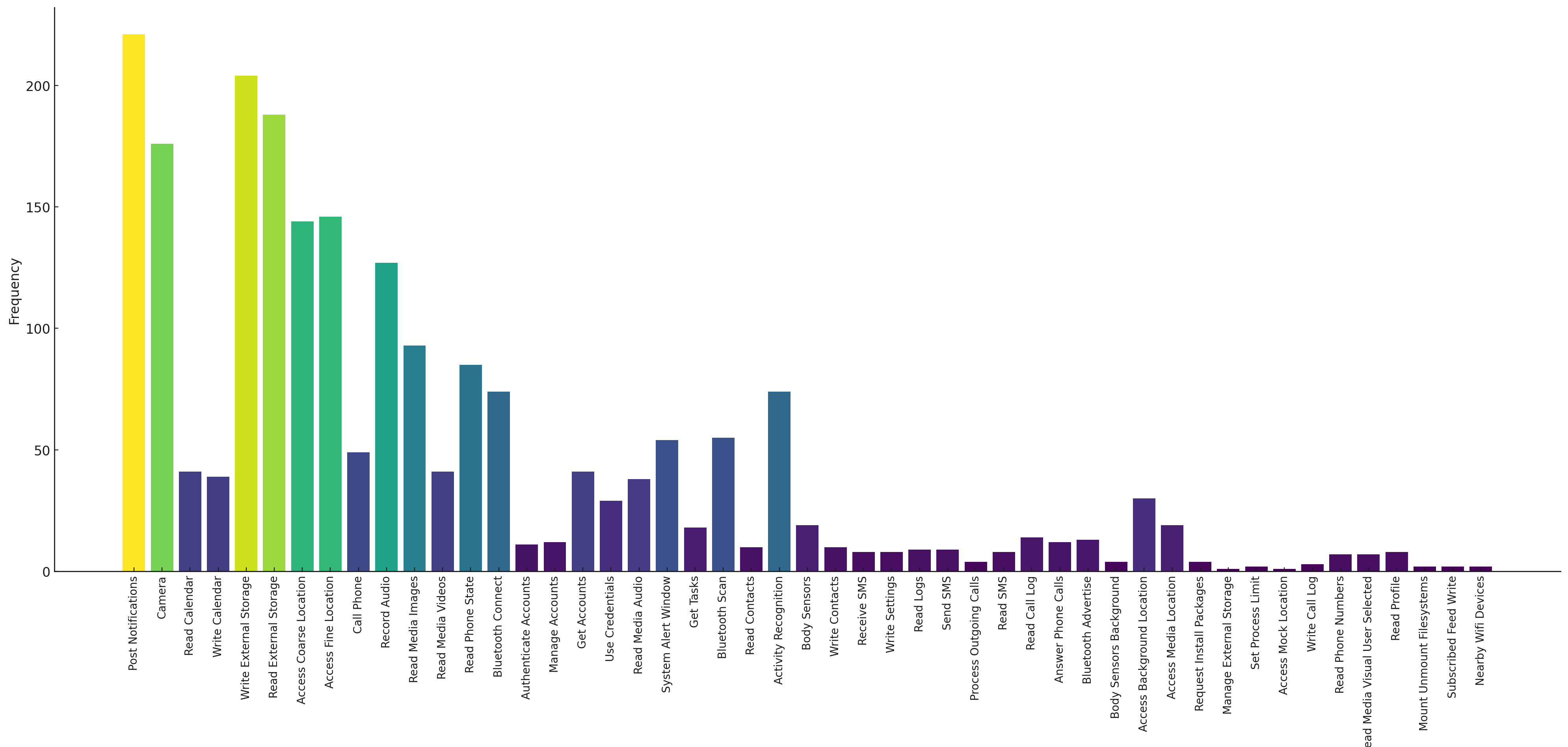}
   \caption{Distribution of dangerous permissions identified across healthcare applications using RiskInDroid.}
    \label{fig:dangerousrisk}
\end{figure*}

RiskInDroid evaluates applications based on permission usage, where a lower score indicates stronger security posture. The most critical category, \emph{Not Required but Used}, represents permissions actively used without being declared to the user, creating a significant transparency and compliance risk. Figure~\ref{fig:dangerousrisk} shows the distribution of these permissions across the analyzed mHealth applications. Across the $150$ applications processed, the average RiskInDroid score was $21.57$, with a minimum of $5.46$ and a maximum of $69.45$. The highest score was associated with an APK declaring $17$ permissions, $10$ of which were used and $7$ required but unused, while also using $10$ undeclared permissions. High-impact undeclared permissions included \texttt{SEND\_SMS}, \texttt{BROADCAST\_STICKY}, \texttt{DISABLE\_KEYGUARD}, \texttt{READ\_PHONE\_STATE}, and \texttt{MANAGE\_ACCOUNTS}. 

Every analyzed application exhibited at least one \emph{Not Required but Used} permission. Key statistics include: $110$ apps using \texttt{AUTHENTICATE\_ACCOUNTS}, $148$ using \texttt{READ\_PROFILE}, $71$ accessing \texttt{ACCESS\_FINE\_LOCATION}, $75$ accessing \texttt{ACCESS\_COARSE\_LOCATION}, $130$ writing to settings via \texttt{WRITE\_SETTINGS}, and $73$ sending SMS messages without declaration. In contrast, the most consistently declared permissions included \texttt{RECEIVE\_BOOT\_COMPLETED} ($171$ apps), \texttt{WAKE\_LOCK} ($296$), \texttt{ACCESS\_WIFI\_STATE} ($145$), and \texttt{INTERNET} ($302$). Table~\ref{table:permission} lists all detected undeclared but used permissions, several of which enable sensitive capabilities such as account manipulation, location tracking, and network state modification. The high prevalence of such permissions underscores the role of embedded SDKs and inherited code in introducing stealthy data access paths outside Android's normal auditing model.

\begin{table}[ht]
\caption{Permissions that are Not Declared but Used by the Apps - RiskInDroid}
\centering
\begin{tabular}{||c c||} 
 \hline
 Permission & Count\\ [0.5ex] 
 \hline\hline
 Send SMS & 73\\
 \hline
 Get Accounts & 104\\
 \hline
 Authenticate Accounts & 110\\
 \hline
 Vibrate & 13 \\
 \hline
Read Profile & 148\\
\hline
Receive Boot Completed & 22\\
\hline
Broadcast Sticky & 77\\
\hline
Access Fine Location & 71\\
\hline
Access Coarse Location & 75\\
\hline
Change WiFi State & 74\\
\hline
Access WiFi State & 18\\
\hline
Modify Audio Settings & 22\\
\hline
Read Phone State & 66\\
\hline
Write Settings & 130\\
\hline
Manage Accounts & 117\\
\hline
Use Credentials & 44\\
\hline
Access Mock Location & 1\\
\hline
Restart Packages & 2\\
\hline
Bluetooth Admin & 2\\
\hline
Get Tasks & 30\\
\hline
Bluetooth & 14\\
\hline
Record Audio & 4\\
\hline
Wake Lock & 3\\
\hline
Disable Keyguard & 1\\
\hline
Change WiFi Multicast State & 2\\[0.5ex] 
 \hline
\end{tabular}
\label{table:permission}
\end{table}

\subsection{OWASP Mobile Audit}
The OWASP Mobile Audit applies static application security testing (SAST) aligned with the OWASP Mobile Application Security Testing Guide (MASTG) to identify vulnerabilities, insecure configurations, certificate handling issues, and potential malware signatures. Across the $95$ mHealth applications analyzed, the mean vulnerability counts per app were: $44$ \emph{critical}, $2,194$ \emph{high}, $3,728$ \emph{medium}, $1,735$ \emph{low}, and $1,469$ with no severity assigned. The most vulnerable case recorded $299$ critical findings, while the highest number of high-severity issues in a single app reached $11,345$. To contextualize these figures, we cross-referenced OWASP's categorization with the OWASP Mobile Top 10 risk taxonomy, which includes: improper credential usage, inadequate supply chain security, insecure authentication, insufficient input/output validation, insecure communication, inadequate privacy controls, insufficient binary protections, security misconfiguration, insecure data storage, and insufficient cryptography. The average of $44$ critical instances per application highlights systemic exposure to these top-tier risks. Frequent critical findings included insecure storage of authentication tokens, absence of certificate pinning, unencrypted PHI transmission, and reliance on outdated cryptographic primitives. The presence of such vulnerabilities at scale suggests that many mHealth applications are deployed without undergoing security verification against recognized industry baselines, leaving both patient data and application integrity susceptible to compromise.

\begin{table*}[htp]
\centering
\caption{Summary of Security and Privacy Findings by Analysis Tool}
\label{tab:tool_summary}
\begin{tabular}{|p{3.0cm}|p{3cm}|p{6cm}|}
\hline
\textbf{Tool} & \textbf{Type of Analysis} & \textbf{Key Findings} \\
\hline
\textbf{MobSF} (272 apps) & Static code and permission analysis & 
134 apps use SHA1 vulnerable to Janus \newline
42 apps permit unencrypted traffic \newline
2 apps bypass certificate pinning \newline
181 apps support insecure OS versions \newline
529 unknown permissions detected \\
\hline
\textbf{RiskInDroid} (150 apps) & Permission classification (declared vs undeclared) & 
73 apps send SMS without disclosure \newline
71 apps access fine location \newline
66 apps read phone state \newline
130 apps write settings silently \\
\hline
\textbf{OWASP Mobile Audit} (95 apps) & Static Application Security Testing (SAST) & 
Avg 44 critical vulnerabilities per app \newline
Top risks: auth, data storage, insecure comms \newline
Max: 299 critical, 11,345 high-severity issues \\
\hline
\end{tabular}
\end{table*}

\subsection{Review Analysis}
We conducted a large-scale sentiment and term-frequency analysis on user feedback for all $272$ applications. Using a Google Play Store scraper, we collected more than $2.5$M reviews and processed them with the TextBlob NLP library for sentiment classification. To reduce noise from bots or low-information feedback, we excluded all reviews with fewer than $100$ characters. After filtering, each application averaged $8,781$ reviews: $6,309$ positive, $1,451$ negative, and $1,020$ neutral. To map user-reported concerns to technical domains, we developed five semantic categories: \emph{confidentiality}, \emph{user information}, \emph{permissions}, \emph{account access}, and \emph{usable security}. Term extraction and classification revealed that \emph{usable security} issues dominated, appearing in $553,495$ reviews. These complaints, consistent with prior observations by Diao et al.~\cite{diao2019kindness}, referenced interface instability, slow performance, non-responsive features, and repeated authentication failures. \emph{Account access} issues appeared in $154,032$ reviews, while \emph{permissions} and \emph{user information} were mentioned in $38,126$ and $38,883$ reviews respectively. \emph{Confidentiality} concerns, including explicit mentions of privacy violations and PHI exposure, occurred in $15,176$ reviews. The category distributions are summarised in Table~\ref{table:reviews}. The results show that usability failures in authentication and data access closely align with security weaknesses, indicating shared engineering flaws that undermine both stability and user trust.

\begin{table}[ht]
\caption{Frequency of the Reviews by Category}
\centering
\begin{tabular}{||c c||} 
 \hline
 Category & Number of Reviews \\ [0.5ex] 
 \hline\hline
 Confidentiality & 15,176 \\ 
 \hline
 User Info & 38,883 \\ 
 \hline
 Permissions & 38,126 \\ 
 \hline
 Account Access & 154,032 \\ 
 \hline
 Usable Security & 553,495 \\ [1ex] 
 \hline
\end{tabular}
\label{table:reviews}
\end{table}

\textbf{Usability}  
A substantial share of user feedback identified persistent failures in authentication workflows, exception handling, and interface stability as barriers to using core functionality. Numerous reviewers reported being locked out of mobile services despite valid credentials, pointing to weaknesses in session lifecycle management, credential verification logic, and consistency between mobile and web authentication flows. These failures were often exacerbated by fragmented password policy enforcement, such as restrictive maximum length limits on mobile clients that conflicted with more permissive web implementations, creating unnecessary friction for security-conscious users. Collectively, such deficiencies were mapped to the \emph{usable security} dimension of our taxonomy, accounting for $553,495$ reviews ($21.6\%$ of all collected reviews) and representing the single most dominant theme of user dissatisfaction across the dataset.
\begin{quote}\itshape
     ``I have tried using this app after uploading to an iPhone and an Android, numerous times, but unfortunately it will not let me log in, says password is wrong, then when I try to reset password it will not allow me... very disappointed as app does not work."
\end{quote}
\begin{quote}\itshape
    ``App limits the password length shorter than the website so it is impossible to log in to the app with a secure password."
\end{quote}
\begin{quote}\itshape
     ``Won't let me log in, keeps saying bad request or user not found even though I have done all the steps and have managed to log in through the website."
\end{quote}

When authentication succeeded, users often faced degraded in-app performance, including non-responsive UI elements, random freezes, and failed core functions without diagnostic feedback. These suggest poor state management, inadequate input validation, and weak event-handling resilience. Users also described the mobile app as less stable than the web client, indicating issues such as improper API error propagation, thread-management flaws, and unoptimized caching.
\begin{quote}\itshape
    ``It's a great app for keeping in touch with your doctor. But, there are usability and user-friendliness problems. Certain functions just freeze without any explanation. You are instructed to call support but there are endless recorded messages with no answer."
\end{quote}
\begin{quote}\itshape
    ``It's nice having access to records from two of my healthcare providers. The overall usability and responsiveness of the phone app are not really great, but that is the case for many health information apps."
\end{quote}

These weaknesses extend beyond usability. Apps with poor user experience often exhibit critical security flaws such as insecure cryptography (\( r = 0.54, p < 0.001 \)) and unencrypted transmissions (\( r = 0.47, p < 0.001 \)), indicating that weak engineering practices drive both functional instability and exploitable vulnerabilities.

\section{Discussion}
\label{sec:discussion}

\subsection*{RQ1: What proportion of mHealth applications exhibit permission overreach, insecure cryptographic implementations, or exploitable vulnerabilities, and how are these weaknesses characterised in static analysis?}
Our analysis shows that security weaknesses in mHealth applications are not isolated defects but systemic characteristics of the ecosystem. Static analysis revealed that $26.1\%$ of applications accessed fine-grained location data without user disclosure, $26.8\%$ initiated calls or sent SMS messages without explicit consent, and $24.3\%$ queried phone state without functional justification. RiskInDroid identified $529$ proprietary or undocumented permissions, most introduced through third-party SDKs, which bypass Android's public permission framework and create opaque capabilities invisible to both users and external auditors. These permissions often originate from embedded advertising or analytics components and can facilitate device fingerprinting, behavioral profiling, and silent data exfiltration.

On the cryptographic front, $49.3\%$ of applications still relied on \texttt{RSA~SHA-1} signatures, exposing them to collision-based exploits such as \emph{Janus}, which permits undetected APK modification and repackaging. Forty-two applications transmitted PHI in cleartext over HTTP, enabling trivial interception on public or enterprise Wi-Fi networks using off-the-shelf packet capture tools. Twenty-two accepted all TLS certificates without validation, and two explicitly disabled certificate pinning, giving adversaries an unchallenged path for man-in-the-middle interception of credentials and medical data. Vulnerability scanning also detected six active \emph{StrandHogg~2.0} cases, enabling task hijacking and privilege escalation without root access, an attack vector well suited for targeted compromise of telemedicine sessions and insurance workflows. These findings reveal an ecosystem in which insecure defaults, unpatched third-party libraries, and opaque dependencies persist across production codebases, ensuring that exploitable conditions remain available to adversaries at scale.

\subsection*{RQ2: To what extent do technical weaknesses such as insecure permission use, outdated cryptography, and misconfigured network security correlate with user-reported privacy, security, and usability issues in mHealth applications?}
Our sentiment analysis of $2.56$M Google Play reviews found that $28.5\%$ carried negative or neutral sentiment ($423,729$ negative, $297,976$ neutral), with $553,495$ explicitly referencing privacy intrusions, data misuse, or operational instability. These perceptions strongly aligned with the vulnerabilities identified: permission overreach correlated with negative or neutral sentiment ($r = 0.62, p < 0.001$), insecure cryptographic implementations ($r = 0.54, p < 0.001$), and unencrypted transmissions ({$r = 0.47, p < 0.001$}). Users frequently reported unjustified access to location, camera, or contacts, mirroring our detection of undeclared or high-risk permissions in static analysis. 

Performance complaints such as persistent crashes, input lag, and long load times often appeared alongside insecure cryptography or permissive network configurations. This pattern suggests that weak engineering practice manifests in both degraded functional reliability and a weakened security posture. Notably, a subset of users explicitly equated instability with insecure handling of personal data, demonstrating that perceived operational quality forms part of an implicit trust model. Once this trust is undermined, reputational damage can occur even in the absence of a confirmed breach, increasing the likelihood of user attrition and regulatory attention.

\subsection*{RQ3: What misconfiguration and permission misuse patterns can be extracted from this evidence, and how can they inform enforceable secure-by-design and regulatory practices?}
Three systemic patterns emerged. First, permission overreach often results from monetization-driven SDK integrations that silently introduce location tracking, device fingerprinting, and behavioral analytics without operational need. Second, cryptographic negligence persists due to the lack of enforced standards for algorithm strength, certificate validation, and PHI encryption in consumer health apps. Third, network misconfigurations such as all-certificate trust and unencrypted traffic reflect reliance on insecure default settings rather than deliberate developer intent.

Addressing these weaknesses requires mandatory secure-by-design enforcement at the tooling and governance levels. Secure defaults should be embedded into SDKs, IDE templates, and CI/CD pipelines so that least-privilege permissions, certificate pinning, and strong cryptographic suites (\texttt{RSA~2048+} with \texttt{SHA-256} or \texttt{ECDSA~P-256}) are automatically configured. App marketplaces should require machine-verifiable pre-publication audits covering permission declarations, cryptographic configurations, and network protections, with automatic rejection of binaries that fail baseline checks. Proprietary permissions must be disclosed, documented, and justified to the same standard as Android's core permission set. Undisclosed capabilities should be treated as explicit policy violations subject to financial penalties and removal from distribution channels. Without such measures, the structural defects we identify will persist and continue to undermine both user trust and the confidentiality of PHI.

\section{Implications}
\label{sec:implications}
Our findings reveal systemic weaknesses in the mHealth ecosystem, which operates at the intersection of sensitive data handling, platform controls, and user trust. Despite regulatory attention, enforcement and secure engineering remain inconsistent. We identify four priority areas encompassing platform enforcement, secure-by-design development, user transparency, and regulatory modernization, all supported by automated socio-technical vetting pipelines.

\subsection{Platform Policy Enforcement}
In this study, we examined Android applications, whose permission framework is designed for least-privilege use, yet enforcement is uneven. Many apps in our dataset bypass consent by exploiting undeclared or proprietary permissions, often accessing sensitive data such as location, SMS, and device identifiers. Google Play's review process prioritizes malware detection but rarely flags static privacy violations, including insecure certificate configurations and privileged API use~\cite{muhammad2023circumventing,andow2019policylint}.  
To close this gap, marketplaces should embed automated tools such as RiskInDroid~\cite{merlo2017riskindroid}, MobSF~\cite{bergadano2020modular}, and OWASP Mobile Audit~\cite{acharya2015owasp} directly into submission workflows. Extending these checks to silent permission use, legacy cryptography, and telephony misuse, combined with manifest-to-behavior audits and update-time differential analysis, would help detect and prevent privilege escalation.

\subsection{By-Design Engineering}
Weak enforcement is compounded by insecure development practices. Nearly half of evaluated apps still rely on deprecated cryptography (SHA-1, MD5), and known exploits such as Janus and StrandHogg 2.0 remain exploitable. This reflects inconsistent integration of SAST/DAST tools into CI/CD pipelines. Developers should adopt the OWASP MASVS~\cite{holguera2022owasp}, embed security linters into IDEs, and align authentication with NIST SP 800-63B~\cite{chatzipavlou2016recommended}. To address risks from third-party SDKs, SBOM generation and automated dependency scanning should become prerequisites for app store submission~\cite{stoddard2023software}.

Users notice when app behavior conflicts with stated purpose. Permission-function mismatches, especially for location, SMS, and contacts, frequently erode trust~\cite{lupton2012m}. Progressive, context-aware permission flows~\cite{bai2010context} can reduce suspicion, while platform rules should prohibit dark patterns that coerce consent, in line with EU Digital Services Act requirements~\cite{heldt2022eu}. At scale, NLP-based monitoring of app reviews could surface privacy concerns early, allowing developers to intervene before reputational damage occurs.

\subsection{Regulation of Gray-Zone Health Apps}
Trust also depends on regulatory coverage. Frameworks such as HIPAA and GDPR focus on clinical providers, leaving many gray-zone apps (e.g., fitness trackers, symptom checkers) outside formal oversight despite their collection of PHI-equivalent data~\cite{istepanian2022mobile}. Regulators should broaden the definition of health data handlers to cover any entity processing biometric or medical information~\cite{carayon2018challenges}. Privacy nutrition labels in app stores and independent certification schemes (e.g., SOC 2–style for mHealth) would provide clearer accountability and verifiable standards~\cite{loser2014security}.

\subsection{Automated Vetting and Socio-Technical Oversight}
Finally, sustaining improvements requires moving beyond manual audits. Continuous vetting pipelines that combine static and dynamic analysis, taint tracking, permission mapping, and traffic anomaly detection can expose covert channels and privilege escalation paths~\cite{nurgalieva2020security}. Sandboxed monitoring of CPU, sensor, and network activity adds runtime assurance. For decision-makers, explainable dashboards built on tools like MobSF and RiskInDroid can translate technical findings into actionable insights~\cite{stoddard2023software}, supporting coordinated socio-technical oversight.

\section{Limitations and Future Work}
\label{sec:limitations}
We conducted an assessment of $272$ mHealth applications, but our analysis faced constraints from tools like \textit{RiskInDroid} and \textit{OWASP Mobile Audit}, which could not process some large APKs due to file size limits. We mitigated this by combining multiple static analysis tools, yet static methods cannot fully capture runtime behaviors such as time-triggered data exfiltration, dynamic code loading, or conditional permission escalation. In future work, we will integrate dynamic analysis capable of monitoring real-time network activity and execution flows, expand coverage to iOS mHealth apps for cross-platform comparison, and perform longitudinal tracking to assess whether identified vulnerabilities are remediated or persist across release cycles.

\section{Conclusion}
\label{sec:conclusion}
mHealth applications are integral to clinical and wellness workflows, yet our analysis of \(272\) Android mHealth apps shows that fundamental security and privacy controls are routinely bypassed. We found \(134\) apps (\(49.3\%\)) still signed with deprecated \texttt{RSA~SHA-1} certificates vulnerable to Janus, \(42\) (\(15.4\%\)) transmitting PHI in cleartext, and \(6\) (\(2.2\%\)) exploitable via \emph{StrandHogg~2.0}. RiskInDroid detected \(529\) undeclared permissions, including \(73\) apps (\(26.8\%\)) silently sending SMS, \(144\) (\(52.9\%\)) accessing fine location without disclosure, and \(66\) (\(24.3\%\)) unnecessarily reading phone state. Additionally, \(22\) apps (\(8.1\%\)) accepted all TLS certificates and \(2\) (\(0.7\%\)) bypassed certificate pinning entirely, creating trivial man-in-the-middle risks. These weaknesses are statistically linked to user distrust: permission overreach (\( r = 0.62, p < 0.001 \)), insecure cryptography (\( r = 0.54, p < 0.001 \)), and unencrypted transmissions (\( r = 0.47, p < 0.001 \)) correlate with \(553{,}495\) user reviews citing privacy breaches, data misuse, or operational instability. Overall, the findings indicate that insecure engineering is a systemic issue within the mHealth ecosystem. Our reproducible methodology, combining static analysis, permission forensics, and sentiment mining, offers a scalable model for app store vetting, supporting integration into review pipelines, mandatory permission disclosure, and stronger regulation for PHI-processing apps.

\section{Acknowledgment}
We thank the Data Agency and Security (DAS) Lab at George Mason University (GMU) and the University of Denver (DU), where this study was conducted. We also thank Dhiman Goswami for helping to edit the final version of this paper. All opinions expressed are solely those of the authors.

\bibliographystyle{IEEEtran}
\bibliography{build}

\end{document}